\documentclass[a4paper,11pt]{article}
\pdfoutput=1 

\usepackage{jcappub} 

\usepackage[T1]{fontenc} 
\usepackage{verbatim}
\usepackage[utf8]{inputenc}
\usepackage[american]{babel}
\usepackage{graphicx}
\usepackage{epsfig}
\usepackage{adjustbox}
\usepackage{booktabs}
\usepackage{multirow}
\usepackage{dcolumn}
\usepackage{amsmath}
\usepackage{amsfonts}
\usepackage{mathtools}
\usepackage{amssymb}
\usepackage{empheq}
\usepackage{subcaption} 
\usepackage{caption}
\usepackage{epstopdf}
\usepackage{bm}
\usepackage{bbm}
\usepackage{dsfont}
\usepackage{mathrsfs}
\usepackage{version}
\usepackage{enumerate}
\usepackage{braket}
\usepackage{enumitem}
\usepackage{transparent}
\usepackage{pifont}
\usepackage{colortbl}
\usepackage{rotating}
\usepackage{adjustbox}
\usepackage{cancel}
\usepackage{tabularx}
\usepackage{soul}
\usepackage{pdfpages}
\usepackage{microtype}
\usepackage{ulem}
\usepackage[table]{xcolor}
\usepackage{siunitx}
\usepackage{float} 
\usepackage[nocompress]{cite} 
\usepackage{marginnote}
\usepackage{relsize}
\usepackage{feynmp-auto}

\usepackage{hyperref}
\hypersetup{colorlinks=true,
linkcolor=navyblue,citecolor=coralred,urlcolor=green(munsell),pdfencoding=auto}

\definecolor{navyblue}{rgb}{0.0, 0.0, 0.5}
\definecolor{bleudefrance}{rgb}{0.19, 0.55, 0.91}
\definecolor{coralred}{rgb}{1.0, 0.25, 0.25}
\definecolor{royalblue}{rgb}{0.25, 0.41, 0.88}
\definecolor{cadmiumgreen}{rgb}{0.0, 0.42, 0.24}
\definecolor{green(munsell)}{rgb}{0.0, 0.66, 0.47}
\definecolor{blue-violet}{rgb}{0.54, 0.17, 0.89}
\definecolor{darkviolet}{rgb}{0.58, 0.0, 0.83}
\definecolor{orange(colorwheel)}{rgb}{1.0, 0.5, 0.0}
\definecolor{internationalorange}{rgb}{1.0, 0.31, 0.0}

\definecolor{magenta(process)}{rgb}{1.0, 0.0, 0.56}

\definecolor{darkspringgreen}{rgb}{0.09, 0.45, 0.27}

\definecolor{royalblue(web)}{rgb}{0.25, 0.41, 0.88}

\definecolor{cadmiumorange}{rgb}{0.93, 0.53, 0.18}

\definecolor{heliotrope}{rgb}{0.87, 0.45, 1.0}

\makeatletter
\renewcommand*{\@textcolor}[3]{%
\protect\leavevmode
\begingroup
\color#1{#2}#3%
\endgroup
}
\makeatother

\newcommand{\myfloatalign}{\centering}

\makeatletter
\newlength{\apb@width}
\newcommand{\autoparbox}[2][c]{\settowidth{\apb@width}{#2}\parbox[#1]{\apb@width}{#2}}

\makeatother

\DeclareCaptionLabelSeparator{colquad}{.\,\,\,}
\DeclarePairedDelimiter{\abs}{\lvert}{\rvert}

\makeatletter
\let\save@mathaccent\mathaccent
\newcommand*\if@single[3]{%
\setbox0\hbox{${\mathaccent"0362{#1}}^H$}%
\setbox2\hbox{${\mathaccent"0362{\kern0pt#1}}^H$}%
\ifdim\ht0=\ht2 #3\else #2\fi
}
\newcommand*\rel@kern[1]{\kern#1\dimexpr\macc@kerna}
\newcommand*\widebar[1]{\@ifnextchar^{{\wide@bar{#1}{0}}}{\wide@bar{#1}{1}}}
\newcommand*\wide@bar[2]{\if@single{#1}{\wide@bar@{#1}{#2}{1}}{\wide@bar@{#1}{#2}{2}}}
\newcommand*\wide@bar@[3]{%
\begingroup
\def\mathaccent##1##2{%
\let\mathaccent\save@mathaccent
\if#32 \let\macc@nucleus\first@char \fi
\setbox\z@\hbox{$\macc@style{\macc@nucleus}_{}$}%
\setbox\tw@\hbox{$\macc@style{\macc@nucleus}{}_{}$}%
\dimen@\wd\tw@
\advance\dimen@-\wd\z@
\divide\dimen@ 3
\@tempdima\wd\tw@
\advance\@tempdima-\scriptspace
\divide\@tempdima 10
\advance\dimen@-\@tempdima
\ifdim\dimen@>\z@ \dimen@0pt\fi
\rel@kern{0.6}\kern-\dimen@
\if#31
\overline{\rel@kern{-0.6}\kern\dimen@\macc@nucleus\rel@kern{0.4}\kern\dimen@}%
\advance\dimen@0.4\dimexpr\macc@kerna
\let\final@kern#2%
\ifdim\dimen@<\z@ \let\final@kern1\fi
\if\final@kern1 \kern-\dimen@\fi
\else
\overline{\rel@kern{-0.6}\kern\dimen@#1}%
\fi
}%
\macc@depth\@ne
\let\math@bgroup\@empty \let\math@egroup\macc@set@skewchar
\mathsurround\z@ \frozen@everymath{\mathgroup\macc@group\relax}%
\macc@set@skewchar\relax
\let\mathaccentV\macc@nested@a
\if#31
\macc@nested@a\relax111{#1}%
\else
\def\gobble@till@marker##1\endmarker{}%
\futurelet\first@char\gobble@till@marker#1\endmarker
\ifcat\noexpand\first@char A\else
\def\first@char{}%
\fi
\macc@nested@a\relax111{\first@char}%
\fi
\endgroup
}
\makeatother

\newcommand{\expect}[1]{\left\langle #1 \right\rangle}

\newcommand\ee{\end{equation}}
\newcommand\be{\begin{equation}}
\newcommand\eea{\end{eqnarray}}
\newcommand\bea{\begin{eqnarray}}
\newcommand{\bsp}{\begin{split}}
\newcommand{\esp}{\end{split}}
\newcommand{\bit}{\begin{itemize}[leftmargin=*]}
\newcommand{\eit}{\end{itemize}}
\newcommand{\ben}{\begin{enumerate}[leftmargin=*]}
\newcommand{\een}{\end{enumerate}}

\renewcommand{\emph}{\textit}

\newcommand\eq[1]{Eq.~\eqref{eq:#1}}

\newcommand{\eqsII}[2]{Eqs.~\eqref{eq:#1}, \eqref{eq:#2}}
\newcommand{\eqsIII}[3]{Eqs.~\eqref{eq:#1}, \eqref{eq:#2}, \eqref{eq:#3}}

\newcommand{\iu}{\mathrm{i}}
\newcommand{\eu}{\mathrm{e}}
\newcommand{\dif}{\mathrm{d}}

\renewcommand{\vec}{\bm}

\newcommand\eps{\varepsilon}

\newcommand{\vk}{\vec{k}}

\def\O{\mathcal{O}}

\def\d{\delta}

\def\dirac{\delta^{(3)}_{\mathrm{D}}}
\def\vx{\vec{x}}
\def\vk{\vec{k}}

\def\<{\left\langle}
\def\>{\right\rangle}

\def\comment#1{}

\title{The Likelihood for LSS: Stochasticity of Bias Coefficients at All Orders}

\author[a]{Giovanni Cabass,}
\author[a]{Fabian Schmidt}

\affiliation[a]{Max-Planck-Institut f\"{u}r Astrophysik, 
Karl-Schwarzschild-Str. 1, 85741 Garching, Germany}

\emailAdd{gcabass@mpa-garching.mpg.de}
\emailAdd{fabians@mpa-garching.mpg.de}

\abstract{\noindent In the EFT of biased tracers the noise field 
$\varepsilon_g$ is not exactly uncorrelated with the nonlinear 
matter field $\delta$. Its correlation with $\delta$ is effectively 
captured by adding stochasticities to each bias coefficient. We show that if these stochastic fields are Gaussian 
(the impact of their non-Gaussianity being subleading on quasi-linear scales anyway) it is possible to resum exactly their effect 
on the conditional likelihood ${\cal P}[\delta_g|\delta]$ to observe a galaxy field $\delta_g$ given an underlying $\delta$. 
This resummation allows to take them into account in EFT-based approaches to Bayesian forward modeling. 
We stress that the resulting corrections to a purely Gaussian conditional likelihood 
with white-noise covariance are the most relevant on scales where the EFT is under control: 
they are more important than {\it any} non-Gaussianity of the noise $\varepsilon_g$.}

\begin{document}
\maketitle
\flushbottom



\section{Introduction}
\label{sec:introduction}

\noindent The effective field theory (EFT) of large-scale structure (LSS) allows for a rigorous, controlled 
incorporation of the effects of fully nonlinear structure formation 
on small scales in the framework of cosmological perturbation theory 
\cite{Baumann:2010tm,Carrasco:2012cv}. This is especially important when 
attempting to infer cosmological information from observed biased tracers 
such as galaxies, quasars, galaxy clusters, the Lyman-$\alpha$ forest, and 
others (see \cite{Desjacques:2016bnm} for a review; in the following, we 
will always refer to the tracers as ``galaxies'' for simplicity): since we currently have no way of simulating the formation of 
galaxies ab initio to nearly the required accuracy, approaches 
which rigorously abstract from this imperfect knowledge of the small-scale 
processes involved in the formation of observed galaxies are highly 
valuable. The prediction for the galaxy density field $\delta_g(\vec{x},\tau) = n_g(\vec{x},\tau)/\bar n_g(\tau)-1$ can be broken 
into two parts: a ``deterministic'' part $\delta_{g,{\rm det}}$ which captures 
the modulation of the galaxy density by long-wavelength perturbations; and 
a stochastic residual which fluctuates due to the stochastic small-scale 
initial conditions. When integrating out small-scale modes, this effectively 
leads to a noise in the galaxy density. 

So far, the calculation of galaxy clustering observables in the EFT has 
largely been restricted to correlation functions, such as the power spectrum and bispectrum. 
Recently, Ref.~\cite{Cabass:2019lqx} presented a derivation of the likelihood of the 
entire galaxy density field $\delta_g(\vec{x},\tau)$ given the nonlinear, 
evolved matter density field, in the context of the EFT. This result offers several advantages 
over previous approaches restricted to correlation functions: 
\begin{itemize}[leftmargin=*] 
\item It puts the deterministic bias expansion of the galaxy density and the stochasticity 
of galaxies on the same footing, clarifying the significance of the latter. 
\item It does not rely on a perturbative expansion of the matter density field. Rather, the 
likelihood is given in terms of the fully nonlinear density field, which can be predicted for example using 
N-body simulations, and thus isolates the truly uncertain aspects of the observed galaxy density. 
\item The likelihood is given by the functional Fourier transform of the generating functional. 
Since the latter generates correlation functions, the derivation of \cite{Cabass:2019lqx} provides a 
correspondence between different terms in the likelihood and correlation functions. 
\item The conditional likelihood of the galaxy density field given the evolved matter density 
field is precisely the key ingredient required in full Bayesian (``forward-modeling'') inference approaches 
\cite{1995MNRAS.272..885F,2010MNRAS.406...60J,2010MNRAS.409..355J,2013MNRAS.432..894J,Wang:2014hia,2015MNRAS.446.4250A}, 
and can be employed there directly \cite{Schmidt:2018bkr,Elsner:2019rql} 
(see \cite{1989ApJ...336L...5B,Schmittfull:2017uhh,Seljak:2017rmr,Modi:2019hnu} for related approaches). 
\end{itemize} 

The likelihood presented in \cite{Cabass:2019lqx} includes the deterministic 
bias relation $\delta_g = \delta_{g,\rm det}[\delta]$ at all orders in perturbations. At leading order, the noise 
follows a multivariate Gaussian distribution with scale-independent and spatially-uniform covariance. 
Ref.~\cite{Cabass:2019lqx} identified the most important correction to this noise covariance as 
being the modulation of the noise amplitude by large-scale density perturbations. 
That is, the ``field-dependent noise covariance'' (or simply ``field-dependent covariance'', as we will call it here), 
was shown to be more relevant than the non-Gaussianity of the noise or its nonlocality 
(captured by higher-derivative terms in the noise covariance). 
In \cite{Cabass:2019lqx} the contributions from the field-dependent covariance were studied perturbatively, 
leading to an Edgeworth-like expansion of the conditional likelihood. 

In this paper, we show that this correction can be included at all orders in perturbations 
(while we still stop at leading order in the derivative expansion), and it can also be generalized to 
take into account the modulation of the noise by other long-wavelength operators. 
Together with the deterministic bias relation mentioned above, we thus have resummed 
the two leading effects in the EFT likelihood of biased tracers. 

Apart from extending the perturbative reach of the likelihood, this resummation 
also offers key advantages for its numerical implementation. Since we show that 
the modulation of the noise by the matter field maintains the Gaussian form of the likelihood, but modifies its 
covariance, we can now begin to include these corrections in the framework presented in 
\cite{Schmidt:2018bkr,Elsner:2019rql}. This would not be possible with an Edgeworth expansion, 
which leads to a likelihood that is neither positive-definite nor normalizable. 

The outline of the paper is as follows. In Section~\ref{sec:review} we review 
the results for the EFT likelihood of \cite{Cabass:2019lqx}, also summarizing the notation that is used in the rest of the paper. 
Our main result is derived in Section~\ref{sec:main_result}. Section~\ref{subsec:renormalization} explains 
how one should interpret our result in light of the process of renormalization in the EFT. 
Section~\ref{subsec:perturbative} then shows how to connect to the perturbative treatment of 
\cite{Cabass:2019lqx}, while Section~\ref{subsec:higher_derivative} discusses how one can perturbatively
include higher-derivative corrections. Finally, in Sections~\ref{subsec:numerical_and_renormalization} and \ref{subsec:marg} 
we look in more detail at how the numerical implementation would proceed.

\section{Review of EFT likelihood without stochasticity of bias coefficients}
\label{sec:review}

\noindent In this section we review the results of \cite{Cabass:2019lqx}, where the EFT likelihood was derived 
under the assumption of Gaussian noise and no stochasticity of the bias coefficients. 
We also take advantage of this section (together with Section~\ref{subsec:renormalization}) to 
explain in detail the origin of the various cutoffs, originally introduced in \cite{Schmidt:2018bkr} and 
then perfected in \cite{2020arXiv200406707S}, that are employed when the EFT likelihood is used 
in the forward modeling framework. 

If we define the galaxy field as $\delta_g$ and the nonlinear matter field as $\delta$, we can write the deterministic bias relation as 
\begin{equation}
\label{eq:review-1}
\delta_g(\vec{x}) = \delta_{g,{\rm det}}[\delta](\vec{x})\,\,,
\end{equation}
where the functional $\delta_{g,{\rm det}}[\delta]$ contains all the operators constructed from the nonlinear matter field $\delta$. 
That is, it gives the deterministic bias expansion. Let us write it as 
\begin{equation} 
\label{eq:main-06}
\delta_{g,{\rm det}}[\delta] = \sum_{O}b_{O}\,{O}[\delta]\,\,. 
\end{equation} 
Here we use the basis of \cite{Mirbabayi:2014zca} (see also Sections~2.2--2.5 of \cite{Desjacques:2016bnm}, 
and see \cite{Senatore:2014eva} for an alternative basis) to write the bias expansion at a fixed time. 
Then, in real space and up to second order in perturbations (and leading order in derivatives) we have 
\begin{equation}
\label{eq:main-05}
\delta_{g,{\rm det}}[\delta](\vec{x}) = b_1\delta(\vec{x}) + \frac{b_2}{2}\delta^2(\vec{x}) + b_{K^2}K^2[\delta]\,\,, 
\end{equation} 
where $K^2 = K_{ij}K^{ij}$ and the tidal field $K_{ij}[\delta]$ is equal to $(\partial_i\partial_j/\nabla^2 - \delta_{ij}/3)\delta$. 

The difference between $\delta_g$ and $\delta_{g,{\rm det}}[\delta]$ that arises from integrating out short-scale modes 
that cannot be described within the EFT is captured by a noise $\eps_g(\vec{x})$. Let us assume that the 
noise is Gaussian with power spectrum $P_{\eps_g}(k)$. Locality (effectively the fact that 
the error we make in describing galaxy clustering via \eqsII{review-1}{main-06} at two different positions 
$\vec{x}_1$ and $\vec{x}_2$ is uncorrelated in the limit of large $\smash{\abs{\vec{x}_1-\vec{x}_2}}$) 
and the absence of preferred directions impose that the noise power spectrum is analytic in $k^2$. I.e.~we have 
\begin{equation}
\label{eq:review-2}
P_{\eps_g}(k) = P^{\{0\}}_{\eps_g} + P^{\{2\}}_{\eps_g}k^2 + \dots\,\,.
\end{equation}
Here the coefficients $\smash{P^{\{n\}}_{\eps_g}}$ have dimensions of a length to the power $n+3$: $\smash{P^{\{0\}}_{\eps_g}}$ 
fixes the size of the noise, while we expect that for $n\geq 2$ we have 
\begin{equation}
\label{eq:review-3}
\frac{P^{\{n\}}_{\eps_g}}{P^{\{0\}}_{\eps_g}}\sim R_\ast^n\,\,, 
\end{equation}
where $R_\ast$ is the typical nonlocality scale of galaxy formation. For dark matter halos, $R_\ast$ is expected to be 
of order of the halo Lagrangian radius $R(M_h)$ or of order of the nonlocality scale for matter $\sim 1/k_{\rm NL}$ (i.e.~the 
scale at which the dimensionless linear matter power spectrum becomes of order one), whichever is larger. 

Let us then take a $\Lambda$ smaller than $1/R_\ast$. We can then split the noise field in a short-wavelength part and a long-wavelength 
one. More precisely, the short-wavelength part is obtained by subtracting 
\begin{equation}
\label{eq:review-4-bis}
\eps_{g,\Lambda}(\vec{k}) = \eps_g(\vec{k})\,\Theta(\Lambda-\abs{\vec{k}}) 
\end{equation}
from $\eps_{g}(\vec{k})$. We are assuming the noise to be Gaussian: therefore, the likelihood for the 
short modes and the long modes factorizes (as does the functional measure ${\cal D}\eps_g$). 
Given that we cannot reliably describe short-wavelength modes, we can just integrate out the short-wavelength 
component of the noise, and remain with a likelihood for $\eps_{g,\Lambda}(\vec{k})$ only. 

What is this likelihood? Since we have chosen $\Lambda$ such that the higher-derivative 
terms of \eq{review-2} are negligible, we can write it as 
\begin{equation}
\label{eq:review-4}
{\cal P}[\eps_g] = \Bigg(\prod_{\abs{\vec{k}}\leq\Lambda}\sqrt{\frac{1}{2\pi P^{\{0\}}_{\eps_g}}}\,\Bigg) 
\exp\Bigg({-\frac{1}{2}}\int_{\vec{k}}\frac{\abs{\eps_{g,\Lambda}(\vec{k})}^2}{P^{\{0\}}_{\eps_g}}\Bigg)\,\,. 
\end{equation} 
The normalization of \eq{review-4} is such that, if $\smash{P^{\{0\}}_{\eps_g}\to 0}$, we recover a Dirac delta 
functional that sets $\eps_{g,\Lambda}$ to zero. 

Let us then multiply this likelihood by a Dirac delta functional 
\begin{equation}
\label{eq:review-5}
\delta^{(\infty)}_{\rm D}\big(\delta_{g,\Lambda}(\vec{k}) - \delta_{g,{\rm det},\Lambda}[\delta_\Lambda](\vec{k}) 
- \eps_{g,\Lambda}(\vec{k})\big)\,\,. 
\end{equation}
Here we have cut both the fields $\smash{\delta_g}$ and $\smash{\delta_{g,{\rm det}}}$ at $\Lambda$, and we have constructed 
the deterministic bias expansion from the matter field after cutting it at $\Lambda$ as well, 
in the same way as it was originally described in \cite{Schmidt:2018bkr}. We will explain in detail 
the origin of these cuts in Section~\ref{subsec:renormalization}. 

If we now functionally integrate over $\eps_{g,\Lambda}$, we obtain the conditional likelihood for 
the galaxy field given the matter field, i.e.~ 
\begin{equation}
\label{eq:review-6}
{\cal P}[\delta_{g,\Lambda}|\delta_\Lambda] = \Bigg(\prod_{\abs{\vec{k}}\leq\Lambda}\sqrt{\frac{1}{2\pi P^{\{0\}}_{\eps_g}}}\,\Bigg) 
\exp\Bigg({-\frac{1}{2}}\int_{\abs{\vec{k}}\leq \Lambda} 
\frac{\abs{\delta_g(\vec{k}) - \delta_{g,{\rm det}}[\delta_\Lambda](\vec{k})}^2}{P^{\{0\}}_{\eps_g}}\Bigg)\,\,. 
\end{equation}
Here we have used the fact that the data and the deterministic galaxy density field have both support for $\abs{\vec{k}}\leq\Lambda$ 
to remove the cutoff from the fields themselves and replace it by a cutoff in the integral $\int_{\vec{k}}$, using the fact that 
these two fields appear quadratically in the likelihood. 

How do the corrections due to the higher-order terms in \eq{review-2} enter? Working in Fourier space it is possible to resum them 
exactly by taking $\smash{P^{\{0\}}_{\eps_g}\to P_{\eps_g}(k)}$ in \eq{review-6} above \cite{Schmidt:2018bkr,Cabass:2019lqx}. 

Let us now make an important step that will be fundamental for the rest of the work. That is, we switch to real space. 
This looks problematic, even if we start from \eq{review-6} (in which the noise power spectrum is constant), because of the 
presence of the cutoff in the integral that selects modes below $\Lambda$. However, thanks again to the fact that both the 
galaxy field and its deterministic expression in terms of $\delta_\Lambda$ appear quadratically in the exponent of \eq{review-6}, 
we can write 
\begin{equation}
\label{eq:review-10-temp}
{\cal P}[\delta_{g,\Lambda}|\delta_\Lambda] = \Bigg(\prod_{\vec{x}}\sqrt{\frac{1}{2\pi P_{\eps_g}^{\{0\}}}}\,\Bigg)\exp\Bigg[
{-\frac{1}{2}}\int\dif^3x\,\frac{\big(\delta_{g,\Lambda}(\vec{x}) - 
\delta_{g,{\rm det},\Lambda}[\delta_\Lambda](\vec{x})\big)^2}{P_{\eps_g}^{\{0\}}}\Bigg]\,\,, 
\end{equation}
where the ``$\Lambda$'' subscripts stand for the fact that: 
\begin{itemize}[leftmargin=*]
\item we cut the field $\delta_g$ in Fourier space and transform it back to real space; 
\item we construct $\delta_{g,{\rm det}}$ from $\delta_\Lambda$, we cut it in Fourier space, and then transform it to real space. 
\end{itemize} 
We then take the difference between $\smash{\delta_{g,\Lambda}(\vec{x})}$ and 
$\smash{\delta_{g,{\rm det},\Lambda}[\delta_\Lambda](\vec{x})}$, square it, and integrate it over all $\vec{x}$. 
Effectively, this tells us that it makes sense to write \eq{review-6} in real space, i.e.~ 
\begin{equation}
\label{eq:review-10}
{\cal P}[\delta_g|\delta] = \Bigg(\prod_{\vec{x}}\sqrt{\frac{1}{2\pi P_{\eps_g}^{\{0\}}}}\,\Bigg)\exp\Bigg[
{-\frac{1}{2}}\int\dif^3x\,\frac{\big(\delta_g(\vec{x})-\delta_{g,{\rm det}}[\delta](\vec{x})\big)^2}{P_{\eps_g}^{\{0\}}}\Bigg]\,\,, 
\end{equation}
if we assume that the fields $\smash{\delta}$ and $\smash{\delta_{g,{\rm det}}[\delta]}$ 
appearing in the integral above are cut at a scale longer than $R_\ast$, c.f.~\eq{review-3}.\footnote{In 
\eqsII{review-10-temp}{review-10} we have, for simplicity, written the overall normalization as a real-space product as well: 
it must be intended as filtered, i.e.~as in \eqsII{review-4}{review-6}.} 
The higher-derivative stochasticities, i.e.~the higher orders in an expansion of the noise power spectrum in $\smash{R_\ast^2k^2}$ 
can be treated perturbatively in real space as long as $\Lambda < 1/R_\ast$ 
(this is the subject of Section~\ref{subsec:higher_derivative}). 

Given \eq{review-10}, the main result of this paper is that the leading corrections to it can be resummed 
by replacing the uniform covariance with one that depends on the matter field, 
\begin{equation}
\label{eq:covariance_replacement}
P_{\eps_g}^{\{0\}}\to P_\eps[\delta](\vec{x})\,\,. 
\end{equation} 
That is, we find 
\begin{equation}
\label{eq:review-11}
{\cal P}[\delta_g|\delta] =
\Bigg(\prod_{\vec{x}}\frac{1}{\sqrt{2\pi P_\eps[\delta](\vec{x})}}\Bigg) 
\exp\Bigg[{-\frac{1}{2}}\int\dif^3x\, 
\frac{\big(\delta_g(\vec{x})-\delta_{g,{\rm det}}[\delta](\vec{x})\big)^2}{P_\eps[\delta](\vec{x})}\Bigg]\,\,. 
\end{equation} 
In the next section we derive this result, together with an expression 
for the field-dependent covariance $\smash{P_\eps[\delta](\vec{x})}$. The functional manipulations that 
we will carry out are effectively in the infinite-$\Lambda$ limit: this is necessary if we want to achieve 
a resummation of the contributions from the stochasticity of bias coefficients. We explain in detail 
in Sections~\ref{subsec:renormalization} and \ref{subsec:perturbative} how the cutoff $\Lambda$ arises, and 
especially how this has to do with the process of renormalization.

\section{Main result}
\label{sec:main_result}

\noindent First, let us compute the conditional likelihood 
including the effect of the stochasticity in the linear bias $b_1$. We will then discuss how to include 
the stochasticities in all bias coefficients. 

Let us consider the bias expansion in real space. If we include only the noise in $b_1$, it reads 
\begin{equation}
\label{eq:pre_main-01}
\delta_g(\vec{x}) = \delta_{g,{\rm det}}[\delta](\vec{x}) + \eps_g(\vec{x}) + \eps_{g,\delta}(\vec{x})\delta(\vec{x})\,\,. 
\end{equation} 
The noise fields $\smash{\eps_g}$ and $\smash{\eps_{g,\delta}}$ are uncorrelated with the matter field. 
If we assume they are Gaussian fields (we will discuss this assumption in more detail in Section~\ref{sec:conclusions}), 
their probability distribution is fully characterized by their covariance ${\rm C}_\eps$ in real or Fourier space. 
In Fourier space and on scales longer than the typical scale of galaxy formation, 
locality and the absence of preferred directions ensure that 
this covariance is diagonal and constant (see Section~2.7 of \cite{Desjacques:2016bnm} 
and the previous section for a discussion), i.e.~ 
\be
\label{eq:CepsF}
\expect{ \eps_i(\vk) \eps_j(\vk') } = ({\rm C}_\eps)_{ij}\, (2\pi)^3 \dirac(\vk+\vk')\,\,.
\ee
Here we have $\{{_i},{_j}\}\in\{{_g},{_{g,\delta}}\}$. In real space this becomes 
\be
\expect{ \eps_i(\vec{x}) \eps_j(\vec{y}) } = ({\rm C}_\eps)_{ij}\, \dirac(\vec{x}-\vec{y})\,\,. 
\ee
Let us assume throughout this section that this actually holds on all scales 
(we will study in detail how to go beyond this assumption in Sections~\ref{subsec:renormalization}, \ref{subsec:perturbative} 
and \ref{subsec:higher_derivative}). With this assumption we can write the joint PDF of $\eps_g,\,\eps_{g,\delta}$ as 
\begin{equation}
\label{eq:pre_main-02}
{\cal P}[\eps_g,\eps_{g,\delta}] = \Bigg(\prod_{\vec{x}}\frac{1}{2\pi\sqrt{\det{\rm C}_\eps}}\Bigg) 
\exp\bigg({-\frac{1}{2}}\int\dif^3x\,\vec{\eps}(\vec{x})\cdot{\rm C}_\eps^{-1}\cdot\vec{\eps}(\vec{x})\bigg)\,\,, 
\end{equation} 
where 
\begin{subequations}
\label{eq:pre_main-03}
\begin{align}
\vec{\eps} &= (\eps_g,\eps_{g,\delta})\,\,, \label{eq:pre_main-03-1} \\
{\rm C}_\eps &= 
\begin{pmatrix}
P^{\{0\}}_{\eps_g} & P^{\{0\}}_{\eps_g\eps_{g,\delta}} \\
P^{\{0\}}_{\eps_g\eps_{g,\delta}} & P^{\{0\}}_{\eps_{g,\delta}} 
\end{pmatrix} 
\,\,. \label{eq:pre_main-03-2}
\end{align}
\end{subequations} 
In \eq{pre_main-03-2} we denoted with a superscript ``$\{0\}$'' the low-$k$ limit of the noise 
auto- and cross-spectra. 

Using \eq{pre_main-01} to rewrite $\eps_g$ in terms of $\smash{\delta_g}$, $\smash{\delta_{g,{\rm det}}[\delta]}$ and 
$\smash{\eps_{g,\delta}}$ we can integrate out the field $\smash{\eps_{g,\delta}}$ with a procedure 
analogous to that followed in \cite{Schmidt:2018bkr}. 
In this way we obtain the conditional likelihood ${\cal P}[\delta_g|\delta]$. More precisely, since all our expressions 
are local in real space, the functional integral reduces to a product of ordinary one-dimensional integrals and we find 
\begin{equation}
\label{eq:pre_main-04}
{\cal P}[\delta_g|\delta] = \Bigg(\prod_{\vec{x}}\frac{1}{\sqrt{2\pi P_\eps[\delta](\vec{x})}}\Bigg)\exp\Bigg[
{-\frac{1}{2}}\int\dif^3x\,\frac{\big(\delta_g(\vec{x})-\delta_{g,{\rm det}}[\delta](\vec{x})\big)^2}{P_\eps[\delta](\vec{x})}\Bigg]\,\,, 
\end{equation} 
where we defined the ``field-dependent covariance'' as 
\begin{equation}
\label{eq:pre_main-05}
P_\eps[\delta](\vec{x}) = P^{\{0\}}_{\eps_g} + 2P^{\{0\}}_{\eps_g\eps_{g,\delta}}\delta(\vec{x}) 
+ P^{\{0\}}_{\eps_{g,\delta}}\delta^2(\vec{x})\,\,. 
\end{equation} 
We see that the likelihood has the same structure as the one of \eq{review-10}, with the only difference being that 
$\smash{P^{\{0\}}_{\eps_g}}$ has been replaced by \eq{pre_main-05}. 

To confirm this result, let us derive it using a different approach, that does not involve integrating out 
noise fields but works at the level of the generating functional for correlation functions. 
First, we know that the conditional likelihood ${\cal P}[\delta_g|\delta]$ is given by the joint likelihood 
${\cal P}[\delta_g,\delta]$ divided by the likelihood ${\cal P}[\delta]$ for $\delta$. 
Then, if we know the form of the generating functional $Z[J_g,J]$, the joint 
likelihood ${\cal P}[\delta_g,\delta]$ can be obtained via its functional Fourier transform over the two currents 
$J_g$ and $J$ as described in \cite{Cabass:2019lqx}. 
Since the generating functional is obtained by integrating over the initial conditions $\delta_{\rm in}$ 
(see \cite{Carroll:2013oxa}, for example), we see that ${\cal P}[\delta_g,\delta]$ is given by a 
functional integral of the following form 
\begin{equation}
\label{eq:main-01}
{\cal P}[\delta_g,\delta] = {\cal N}^2_{\delta^{(\infty)}}\int{\cal D}X_g\,{\cal D}X\,{\cal D}\delta_{\rm in}\,
\eu^{\int_{\vec{x}}\vec{\phi}_g(\vec{x})\cdot\vec{\mathcal{J}}_g(\vec{x})-S_g[\vec{\phi}_g]}\,\,,
\end{equation}
where the factor $\smash{{\cal N}^2_{\delta^{(\infty)}}}$ is the infinite-dimensional generalization of a 
$1/(2\pi)^2$ factor that comes with the functional Fourier transform over both currents ($J_g$ and $J$). 
In \eq{main-01} we have also defined 
\begin{subequations}
\label{eq:main-02}
\begin{align}
\vec{\phi}_g &= (X_g,X,\delta_{\rm in})\,\,, \label{eq:main-02-1} \\ 
\vec{\mathcal{J}}_g &= (\iu\delta_g,\iu\delta,0)\,\,. \label{eq:main-02-2} 
\end{align}
\end{subequations} 
That is, the fields $X_g$ and $X$ are the ``momenta'' dual 
to the galaxy and matter fields, $\delta_g$ and $\delta$, in the functional Fourier transform. 
The ``action'' $S_g$ is the sum of a part quadratic in the fields and higher-order interactions, $S_{g,{\rm int}}$. 
As warm-up, let us start by assuming we only have a Gaussian noise field $\varepsilon_g$ with constant power spectrum. 
That is, we do not yet include the effect of $\eps_{g,\delta}$. This allows us to write down $S_g$ exactly, since 
we do not have terms with powers of $X_g$ higher than two in $\smash{S_{g,\rm int}}$ (cf.~the summary in Tab.~\ref{tab:summary_Z}). 
More precisely, \eq{main-01} becomes 
\begin{equation}
\label{eq:main-03}
\begin{split}
{\cal P}[\delta_g,\delta] &= {\cal N}_{\delta^{(\infty)}}^2\int{\cal D}X_g\,{\cal D}X\,{\cal D}\delta_{\rm in}\,
\eu^{\iu\int_{\vec{x}}X_g(\vec{x})\delta_g(\vec{x})}\,\eu^{\iu\int_{\vec{x}}X(\vec{x})\delta(\vec{x})} \\ 
&\hphantom{= {\cal N}_{\delta^{(\infty)}}^2\int{\cal D}X_g\,{\cal D}X\,{\cal D}\delta_{\rm in} } 
\times{\cal P}[\delta_{\rm in}]\,
\eu^{-\frac{1}{2}\int_{\vec{x}}P_{\eps_g}^{\{0\}}{X_{g}^2(\vec{x})}} \\ 
&\hphantom{= {\cal N}_{\delta^{(\infty)}}^2\int{\cal D}X_g\,{\cal D}X\,{\cal D}\delta_{\rm in} } 
\times\eu^{-\iu\int_{\vec{x}}X_g(\vec{x})\delta_{g,{\rm fwd}}[\delta_{\rm in}](\vec{x})}\, 
\eu^{-\iu\int_{\vec{x}}X(\vec{x})\delta_{\rm fwd}[\delta_{\rm in}](\vec{x})}\,\,, 
\end{split}
\end{equation}
where we see that the higher-order interactions in $\smash{S_{g,\rm int}}$ describe the nonlinear forward models 
for galaxies and the gravitational evolution of the initial matter field. 
These forward models are the functionals $\smash{\delta_{\rm fwd}}$ and $\smash{\delta_{g,{\rm fwd}}}$, respectively. 
We can decompose $\smash{\delta_{g,{\rm fwd}}[\delta_{\rm in}]}$ as 
\begin{equation}
\label{eq:main-04}
\delta_{g,{\rm fwd}}[\delta_{\rm in}] = \delta_{g,{\rm det}}\big[\delta_{{\rm fwd}}[\delta_{\rm in}]\big]\,\,, 
\end{equation} 
where the deterministic bias expansion is defined in \eqsII{main-06}{main-05}. 
This equation will continue to hold even after we include the stochasticity of the bias coefficients, 
since their effect is described by a different kind of interaction term in $\smash{S_{g,\rm int}}$.

\begin{table}
\myfloatalign
\caption[.]{Summary of the terms in the expansion of $\smash{S_{g,{\rm int}}}$ 
in powers of $\smash{X_g}$ and $\smash{\delta_{\rm in}}$, and what they correspond to. 
The terms are organized according to their relevance in the infrared, the most relevant being on top. 
Notice that we assume an exact forward model for matter. Hence we are not considering 
terms with more than one power of $X$ in the action, cf.~\eqsII{main-03}{main-08}.} 
\label{tab:summary_Z}
\centering
\medskip
\begin{tabular}{ll}
\toprule
$S_{g,{\rm int}}\supset{}$ & corresponds to \\
\midrule
$X\delta_{\rm in}\delta_{\rm in}\cdots$ & nonlinear deterministic evolution for $\delta$ \\ 
$X_g\delta_{\rm in}\delta_{\rm in}\cdots$ & nonlinear deterministic evolution for $\delta_g$ \\ 
$X_gX_g\cdots \delta_{\rm in}\cdots$ & stochasticities in bias coefficients for $\delta_g$ \\
$X_gX_gX_g\cdots$ & higher-order $n$-point functions of $\eps_g$ \\
\bottomrule
\end{tabular}
\end{table}

What are these terms? As we anticipated above, the stochasticity of the bias coefficients corresponds to interactions of 
the form $X_gX_g\cdots\delta_{\rm in}\cdots$ \cite{Cabass:2019lqx} (see Tab.~\ref{tab:summary_Z}). 
Most importantly, if we assume that such stochasticities are Gaussian the number of powers of $X_g$ is equal to $2$, 
i.e.~we only have terms of the form $X_gX_g\delta_{\rm in}\cdots$. 
The scaling dimensions of the fields $X_g$ and $\delta_{\rm in}$ can be derived from the quadratic part of the action 
in \eq{main-03}. We have $[X_g] = 3/2$, while for Gaussian initial conditions with a power-law power spectrum $P_{\rm in}\propto k^{n_\delta}$ 
we obtain $[\delta_{\rm in}] = (3+n_\delta)/2$. In our Universe $n_\delta$ is close to $-2$: this tells us that for a given number 
of external legs the non-Gaussianity of the noise is very suppressed with respect to the interactions we are considering here.

A further simplification arises if we stop at leading order in the derivative expansion. 
If we assume that only $\smash{P^{\{0\}}_{\eps_g\eps_{g,{O}}}}$ is non-vanishing it is possible to write down 
exactly the form of the interactions $S_{g,{\rm int}}\supset X_gX_g\delta_{\rm in}\cdots$. Indeed, in \cite{Cabass:2019lqx} 
(see its Appendix D) we have shown that they are obtained by shifting the bias coefficients $b_O$ of \eq{main-06} as 
\begin{equation}
\label{eq:main-07}
b_{O}\to b_{O} - \iu P_{\varepsilon_g\varepsilon_{g,{O}}}^{\{0\}} X_g(\vec{x})
\end{equation}
in real space, where the cross-stochasticity $\smash{P_{\varepsilon_g\varepsilon_{g,{O}}}^{\{0\}}}$ 
is a constant of dimensions of length cubed (as in \eq{pre_main-03-2}, for example). 
Using \eq{main-07}, the expression for the joint likelihood then becomes 
\begin{equation}
\label{eq:main-08}
\begin{split}
{\cal P}[\delta_g,\delta] &= {\cal N}_{\delta^{(\infty)}}^2\int{\cal D}X_g\,{\cal D}X\,{\cal D}\delta_{\rm in}\,
\eu^{\iu\int_{\vec{x}}X_g(\vec{x})\big(\delta_g(\vec{x}) - \delta_{g,{\rm fwd}}[\delta_{\rm in}](\vec{x})\big)}\,
\eu^{\iu\int_{\vec{x}}X(\vec{x})\big(\delta(\vec{x}) - \delta_{\rm fwd}[\delta_{\rm in}](\vec{x})\big)} \\ 
&\hphantom{= {\cal N}_{\delta^{(\infty)}}^2\int{\cal D}X_g\,{\cal D}X\,{\cal D}\delta_{\rm in} } 
\times{\cal P}[\delta_{\rm in}]\,\eu^{-\frac{1}{2}\int_{\vec{x}}P_{\eps}[\delta_{\rm fwd} [\delta_{\rm in}]](\vec{x})
{X_{g}^2(\vec{x})}}\,\,,
\end{split}
\end{equation}
where the field-dependent covariance $\smash{P_{\varepsilon}\big[\delta_{\rm fwd}
[\delta_{\rm in}]\big](\vec{x})}$ is defined similarly to \eq{pre_main-05}, i.e.~ 
\begin{equation}
\label{eq:main-09}
\begin{split}
P_{\varepsilon}\big[\delta_{\rm fwd}[\delta_{\rm in}]\big](\vec{x}) &= P^{\{0\}}_{\eps_g} 
+ 2\sum_{O}P_{\varepsilon_g\varepsilon_{g,{O}}}^{\{0\}} {O}\big[\delta_{\rm fwd} [\delta_{\rm in}]\big](\vec{x})\,\,. 
\end{split}
\end{equation}

Thanks to locality (the field $X_g$ is always evaluated at the same position $\vec{x}$ at the order in derivatives we are 
working at) it is now straightforward to carry out the functional integral in $X_g$, since it is a Gaussian integral. We obtain 
\begin{equation}
\label{eq:main-10}
\begin{split}
{\cal P}[\delta_g,\delta] &= {\cal N}_{\delta^{(\infty)}} 
\int{\cal D}X\,{\cal D}\delta_{\rm in}\,\Bigg(\prod_{\vec{x}}
\frac{1}{\sqrt{2\pi P_{\varepsilon}[\delta_{\rm fwd}[\delta_{\rm in}]](\vec{x})}}\Bigg) \\ 
&\hphantom{= {\cal N}_{\delta^{(\infty)}}^2\int{\cal D}X_g\,{\cal D}X\,{\cal D}\delta_{\rm in} } 
\times{\cal P}[\delta_{\rm in}]\,\eu^{\iu\int_{\vec{x}}X(\vec{x})\big(\delta(\vec{x}) 
- \delta_{\rm fwd}[\delta_{\rm in}](\vec{x})\big)} \\ 
&\hphantom{= {\cal N}_{\delta^{(\infty)}}^2\int{\cal D}X_g\,{\cal D}X\,{\cal D}\delta_{\rm in} } 
\times\exp\Bigg[{-\frac{1}{2}}\int\dif^3x\,\frac{\big(\delta_g(\vec{x})-\delta_{g,{\rm fwd}}[\delta_{\rm in}](\vec{x})\big)^2}
{P_\eps[\delta_{\rm fwd}[\delta_{\rm in}]](\vec{x})}\Bigg]\,\,. 
\end{split}
\end{equation}

Finally, we can carry out the integrals over $X$ and $\delta_{\rm in}$. The integral over $X$ gives a Dirac delta functional 
\begin{equation}
\label{eq:main-12}
{\cal N}_{\delta^{(\infty)}}^{-1}\delta^{(\infty)}\big(\delta-\delta_{\rm fwd}[\delta_{\rm in}]\big)\,\,,
\end{equation}
and integrating over $\delta_{\rm in}$ sets $\delta_{\rm in} = \delta^{-1}_{\rm fwd}[\delta]$. Following the same steps for 
the matter likelihood, it is then possible to recognize in \eq{main-10} the conditional likelihood ${\cal P}[\delta_g|\delta]$: 
\begin{equation}
\label{eq:main-13}
{\cal P}[\delta_g|\delta] =
\frac{{\cal P}[\delta_g,\delta]}{{\cal P}[\delta]} = \Bigg(\prod_{\vec{x}}\frac{1}{\sqrt{2\pi P_\eps[\delta](\vec{x})}}\Bigg) 
\exp\Bigg[{-\frac{1}{2}}\int\dif^3x\, 
\frac{\big(\delta_g(\vec{x})-\delta_{g,{\rm det}}[\delta](\vec{x})\big)^2}{P_\eps[\delta](\vec{x})}\Bigg]\,\,. 
\end{equation} 
Using \eq{main-09} with all the stochasticities of $b_{O}$ set to zero except 
for ${O}=\delta$ we recognize the result of \eqsII{pre_main-04}{pre_main-05}. 

So far we have discussed the case of only $\eps_{g,\delta}$ being different from zero, which led us 
to \eqsII{pre_main-04}{pre_main-05}, and the case of all the noises $\eps_{g,O}$ being non-vanishing but 
considering only the impact of $\smash{P_{\varepsilon_g\varepsilon_{g,{O}}}^{\{0\}}}$, which resulted in \eqsII{main-09}{main-13}. 
Before proceeding let us then briefly discuss what happens if we turn on all $\eps_{g,O}$ but do not put 
$\smash{P^{\{0\}}_{\eps_{g,{O}}\eps_{g,{O}'}}}$ to zero. For example, let us consider the stochasticity 
in $O = \delta^2$. The calculation leading to \eqsII{pre_main-04}{pre_main-05} can be straightforwardly 
extended to accomodate the corresponding stochastic field $\eps_{g,\delta^2}$. 
Thanks to locality we now have to solve a two-dimensional integral at each point $\vec{x}$. The resulting conditional likelihood has 
the same form as before, only with a different field-dependent covariance. Indeed, $P_{\eps}[\delta]$ is now given by 
\begin{equation}
\label{eq:main-14}
P_{\eps}[\delta](\vec{x}) = P^{\{0\}}_{\eps_g} 
+ 2\sum_{O}P_{\varepsilon_g\varepsilon_{g,{O}}}^{\{0\}} 
{O}[\delta](\vec{x}) 
+ \sum_{{O},{O}'}P_{\varepsilon_{g,{O}}\varepsilon_{g,{O}'}}^{\{0\}} 
{O}[\delta](\vec{x}){O}'[\delta](\vec{x})\,\,,
\end{equation} 
with 
\begin{equation}
\label{eq:main-15}
{O},{O}'\in\big\{\delta,\delta^2\big\}\,\,. 
\end{equation} 
Combined with the result of \eq{main-09} this equation \emph{strongly} suggests that once we include 
the stochasticities of all the bias coefficients the field-dependent noise keeps the same form 
as in \eq{main-14}, but with ${O},{O}'$ running over \emph{all} the operators of the deterministic bias expansion.

\section{Discussion and conclusions}
\label{sec:conclusions}

\subsection{Renormalization at the field level}
\label{subsec:renormalization}

\noindent In this section we sketch how the process of renormalization, central to the EFTofLSS, would proceed at the field level 
instead of at the level of correlation functions. While a more detailed discussion will be the focus of future work, 
this section is self-contained and allows to make the manipulations of Section~\ref{sec:review} rigorous 
(and to connect with the perturbative treatment of \cite{Cabass:2019lqx}, to which we devote the next section). 

Let us take the generating functional $Z[J_g,J]$ of correlation functions for the galaxy field and matter field. 
It is given by a functional over the initial conditions, which we can write as 
\begin{equation}
\label{eq:renormalization-1}
\begin{split}
Z[J_g,J] &= \int{\cal D}\delta_{\rm in}\,
{\cal P}[\delta_{\rm in}]\,
\eu^{\int_{\vec{x}}J_g(\vec{x})\delta_{g,{\rm fwd}}[\delta_{\rm in}](\vec{x})}\, 
\eu^{\int_{\vec{x}}J(\vec{x})\delta_{\rm fwd}[\delta_{\rm in}](\vec{x})}\,\,. 
\end{split}
\end{equation} 
Here we see that we have an integral over \emph{all} modes of $\delta_{\rm in}$, and we have not included any noise term, 
only the deterministic evolution for galaxies and matter. Moreover, we have not exactly specified what is the form of 
the currents $J_g$ and $J$. 

Let us start from the latter point. When we derive the generating functional with respect to the currents $J_g$ and $J$, 
these fix the external momenta of our correlation functions. We want to probe the correlators in the long-wavelength regime, 
so we must assume that these currents do not have any support above a cutoff $\Lambda$ which we take to be smaller than 
the physical cutoff of our effective description of galaxy clustering. Hence, we have $J_g = J_{g,\Lambda}$, $J=J_\Lambda$. 

What about the other two points? They are tightly related. Indeed, the integral over all modes of $\delta_{\rm in}$ must 
be renormalized. For example, we split $\delta_{\rm in}$ in a short-wavelength part that has support for $\abs{\vec{k}} > \Lambda$ 
and a long-wavelength one that has support for $\abs{\vec{k}}\leq\Lambda$. We can then carry out the integral over $\delta_{\rm in}$ 
by first integrating over the short modes and then over the long modes. Since the integral over the short modes is UV-sensitive, 
counterterms are needed to renormalize it. These counterterms give rise to a noise term in the galaxy power spectrum 
(a term with two powers of $J_{g,\Lambda}$ in the logarithm of the generating functional), 
and a stochasticity of the bias coefficients.\footnote{This process involves the same loops we 
encounter when we want to renormalize correlation functions.} 

After we have added all these counterterms to make the integral over the short modes well-defined, we can carry out the integral over 
the remaining modes, i.e.~$\delta_{\rm in,\Lambda}$. Actually, we can do more. 
We can take the Fourier transform of the generating functional by integrating over $X_{g,\Lambda} = \iu J_{g,\Lambda}$ 
and $X_\Lambda = \iu J_\Lambda$, see~\eq{main-01}. These path integrals are now all well-defined since they 
involve fields that do not have support on arbitrarily short scales. 

However, since there is now an infinite number of terms in $\smash{S_{g,{\rm int}}}$ besides the deterministic evolution, 
i.e.~all the noise counterterms, these path integrals cannot be done in a closed form. Let us see, then, how we arrive 
at the resummation of \cite{Cabass:2019lqx} for the case of Gaussian noise. Essentially, in that case we assume that 
all the renormalized coefficients for the counterterms besides the one in the constant part of the noise power spectrum 
vanish. This is of course an assumption that is not justified from the point of view of the renormalization group. Even 
if we fix to zero the coefficients of the counterterms for the theory at a scale $\Lambda$, changing this scale slightly 
will make the coefficients run according to the renormalization group equations. However, the important part is 
that the fields appearing in $S_{g,{\rm int}}$ are all long-wavelength fields: while the coefficients of the various operators 
may change a little, the importance of the interactions is governed by the linear matter power spectrum. 

Let us then do the Fourier transform of the generating functional of \eq{main-03}, where now all the fields we are integrating over 
only have support for $\smash{\abs{\vec{k}}\leq\Lambda}$. This integral is the one studied in \cite{Cabass:2019lqx}. The calculation 
goes through in the same way as in that paper, and the result is exactly the conditional likelihood of \eq{review-10-temp}, 
\emph{including} the cuts on the fields $\smash{\delta_g}$, $\smash{\delta_{g,{\rm det}}}$ and $\delta$ at $\Lambda$. 
The reason is because the fields $\smash{\delta_g}$ and $\delta$, together with the functionals 
$\smash{\delta_{g,{\rm det}}[\delta_{g,{\rm fwd}}[\delta_{\rm in}]]}$ and $\smash{\delta_{g,{\rm fwd}}[\delta_{\rm in}]}$, 
are linearly coupled to $X_g$ and $X$ in \eq{main-03}. Since these two fields are cut at $\Lambda$ (we are probing our 
theory on large scales), once we integrate in $\dif^3x$ this cut translates to the other fields and functionals as well. 

We are now in position to discuss what happens if we want to resum the stochasticity in the bias coefficients. For simplicity 
we focus on the stochasticity in $b_1$ and only on the term of \eq{main-07} (with $O=\delta$) in $\smash{S_{g,{\rm int}}}$. 
The only difference with the calculation of Section~\ref{sec:main_result} is that to obtain the resummed likelihood 
for the long-wavelength fields the renormalized coefficients of the counterterms 
must be tuned to zero \emph{after} the manipulations that bring us from \eq{main-08} to \eq{main-10} have 
been performed. More precisely, we can always tune the counterterms in such a way that the end result is still \eq{main-13}, 
but with the fields appearing in it (both in the field-dependent covariance and 
the numerator of the exponential) having support only for $\smash{\abs{\vec{k}}\leq\Lambda}$.\footnote{In the 
case of the stochasticity in $b_1$ we would need to tune, for example, the counterterms of the stochasticities of higher-order 
LIMD bias coefficients. This can be seen by expanding $\smash{1/P_{\varepsilon}[\delta_{\rm fwd} 
[\delta_{\rm in}]](\vec{x})}\sim 1/(1+\delta_{\rm in}(\vec{x}))$ in powers of $\delta_{\rm in}$.} 

In the next section we are going to discuss how this leads to a likelihood that is under perturbative control.

\subsection{Connection to perturbative treatment}
\label{subsec:perturbative}

\noindent Let us first study the structure of the result of \eqsII{pre_main-04}{main-13}. 
The noise auto- and cross-spectra all have dimensions of length cubed. 
Factoring out the power spectrum of $\eps_g$ we have that 
\begin{equation}
\label{eq:conclusions-01}
\frac{1}{P_\eps[\delta](\vec{x})} = \frac{1}{P_{\eps_g}^{\{0\}}}\sum_{n=0}^{+\infty}c_n \delta^n(\vec{x})\,\,,
\end{equation}
where $c_n$ are tracer-dependent dimensionless constants which are expected to be of order unity, 
and we have restricted the set of bias operators to powers of the matter density for simplicity. 

Therefore the logarithm $\wp[\delta_g|\delta] \equiv -2\ln {\cal P}[\delta_g|\delta]$ of the conditional likelihood contains only 
terms of the form (forgetting for a moment about the determinant of the inverse covariance) 
\begin{equation}
\label{eq:conclusions-02-a}
\wp[\delta_g|\delta] = \sum_{n=0}^{+\infty} c_n\int\dif^3x\,\delta^n(\vec{x})\,
\frac{\big(\delta_g(\vec{x})-\delta_{g,{\rm det}}[\delta](\vec{x})\big)^2}{P^{\{0\}}_{\eps_g}}\,\,.
\end{equation}
This had to be expected given the structure of the interaction terms in $S_{g,{\rm int}}$ that describe the 
Gaussian noise of the bias coefficients, and matches with the tree-level calculation carried out in \cite{Cabass:2019lqx}. 

We can further connect with the perturbative calculation of \cite{Cabass:2019lqx} by studying the size of 
these corrections with respect to the Gaussian conditional likelihood with field-independent covariance. 
We see immediately that on quasi-linear scales, where the EFT of 
biased tracers is under control, the additional terms that we obtain in \eq{conclusions-02-a} 
are subleading since we include only modes below some cutoff $\ll k_{\rm NL}$, and consequently 
the typical size of a fluctuation $\delta(\vec{x})$ is smaller than unity (see the previous section for 
a more detailed discussion). 

We can also discuss the relative importance of the terms in \eq{conclusions-02-a} with respect to corrections coming from 
the non-Gaussianities of the noise. Noise non-Gaussianities are captured by interactions with more than two powers of $X_g$ 
in $S_{g,{\rm int}}$ (see Tab.~\ref{tab:summary_Z}). They correspond to terms of higher order in the difference 
$\delta_g - \delta_{g,{\rm det}}[\delta]$ in $\wp[\delta_g|\delta]$. Therefore, in an expansion 
\begin{equation}
\label{eq:conclusions-02-b}
\wp[\delta_g|\delta] = \sum_{m=2}^{+\infty}\sum_{n=0}^{+\infty}d_{m,n} \int\dif^3x\,\delta^n(\vec{x})\,
\frac{\big(\delta_g(\vec{x})-\delta_{g,{\rm det}}[\delta](\vec{x})\big)^m}{P^{\{0\}}_{\eps_g}}\,\,,
\end{equation}
again involving dimensionless coefficients $d_{m,n}$ assumed to be of order unity, 
the terms coming from the stochasticity in the bias coefficients are always more relevant than non-Gaussianities at a fixed $m+n$ 
(i.e.~at a fixed number of external legs in $\smash{S_{g,{\rm int}}}$). Indeed, they are always enhanced by powers of the ratio 
\begin{equation}
\label{eq:conclusions-02-c}
\sqrt{\frac{P_{\rm L}(k)}{P^{\{0\}}_{\eps_g}}}\,\,, 
\end{equation} 
where $P_{\rm L}$ is the linear matter power spectrum. Note that if we compare contributions 
at different $m+n$ it is very much possible for terms coming from noise non-Gaussianities to be more important than the ones we 
are keeping non-perturbatively in the field-dependent covariance.

On the other hand, if we expand also the square of the difference between $\delta_g$ and $\delta_{g,{\rm det}}[\delta]$ 
in \eq{conclusions-02-a} around a linear bias relation we see that including an operator $\smash{O[\delta]}$ in 
$\smash{\delta_{g,{\rm det}}[\delta]}$ is always more important on large scales than 
including the stochasticity in its bias coefficient: the former comes with an enhancement by one power 
of the ratio in \eq{conclusions-02-c} with respect to the latter. 

Finally, let us discuss how to treat the determinant of the field-dependent noise in \eqsII{pre_main-04}{main-13}. 
More precisely, we want to make the connection with one of the results of \cite{Cabass:2019lqx}. There we have shown that 
once the stochasticity of bias coefficients is included, loops of the field $X_g$ generate counterterms in the action that 
carry only powers of the initial field $\delta_{\rm in}$. These new interactions give rise to terms in the 
log-likelihood $\wp[\delta_g|\delta]$ that do not depend on the ``data'' $\delta_g$, but only on the matter field $\delta$: 
i.e.~to terms with $m=0$ in \eq{conclusions-02-b} (which we haven't included there). We can straightforwardly 
see that they correspond exactly to the determinant in \eqsII{pre_main-04}{main-13} by using the relation 
\begin{equation}
\label{eq:conclusions-03}
\prod_{\vec{x}}\frac{1}{\sqrt{2\pi P_\eps[\delta](\vec{x})}} = 
\eu^{-\frac{1}{2}\int_{\vec{x}}\ln 2\pi P_{\varepsilon}[\delta](\vec{x})}\,\,.
\end{equation}

\subsection{Including higher-derivative stochasticity}
\label{subsec:higher_derivative}

\noindent In this paper we have shown how a field-dependent stochasticity can be incorporated 
into the EFT likelihood at all orders in perturbations if we stop at the lowest (zeroth) order in derivatives. 

In addition to this contribution (and the non-Gaussianity of the stochasticity discussed in Section~\ref{subsec:perturbative}), 
however, we also have higher-derivative stochastic terms. These correspond to 
a series in $k^2$ in the Fourier-space covariance of \eq{CepsF}. Refs.~\cite{Schmidt:2018bkr,Elsner:2019rql} 
argued that these contributions can be resummed when writing the likelihood in Fourier space. 

In terms of scaling dimensions, in \cite{Cabass:2019lqx} we have shown that the field-dependent 
stochasticity is more relevant than the higher-derivative stochasticity; 
specifically, relative to the leading (constant) Gaussian stochasticity 
the former is suppressed by $(3+n_\delta)/2 \sim 0.8$, while the latter is suppressed by $2$. 
Since one cannot resum both of these contributions at the same time in closed form (the reason being 
that derivatives are local operations in Fourier space, while multiplications are local operations in real space), 
it thus makes sense to resum the one that is more relevant on large scales, as done here. 

Nevertheless it is possible to incorporate the higher-derivative stochasticity in the result of this work in 
a perturbative way. Let us now show how. First, we extend \eq{CepsF} to 
\be
\label{eq:CepsF2}
\expect{ \eps_i(\vk) \eps_j(\vk') } = \big({\rm C}_\eps + {\rm C}_\eps^{\{2\}} k^2\big)_{ij}\, (2\pi)^3 \dirac(\vk+\vk')\,\,. 
\ee
In real space this becomes
\be
\expect{ \eps_i(\vec{x}) \eps_j(\vec{y}) } = \big({\rm C}_\eps - {\rm C}_\eps^{\{2\}} \nabla^2\big)_{ij}\, \dirac(\vec{x}-\vec{y})\,\,. 
\ee
Then, expanding to leading order in ${\rm C}_\eps^{\{2\}}/{\rm C_\eps}$, 
we can write the joint PDF of $\eps_g,\,\eps_{g,\delta}$ as 
(we neglect the expansion of the determinant in the following, since this section is meant to give only a qualitative discussion) 
\begin{equation} 
{\cal P}[\eps_g,\eps_{g,\delta}] = \Bigg(\prod_{\vec{x}}\frac{1}{2\pi\sqrt{\det{\rm C}_\eps}}\Bigg) 
\exp\bigg({-\frac{1}{2}}\int\dif^3x\,\vec{\eps}(\vec{x})\cdot{\rm C}_\eps^{-1}\cdot 
\left(\mathds{1} + {\rm C}_\eps^{-1}\cdot{\rm C}_\eps^{\{2\}} \nabla^2\right)
\cdot\vec{\eps}(\vec{x})\bigg)\,\,.
\end{equation} 
Let us now only keep the $i=j=1$ entry of the matrix $\smash{\big({\rm C}_\eps^{-1}\cdot{\rm C}_\eps^{\{2\}}\big)_{ij}}$, defining 
\begin{equation}
	R_\eps^2 \propto \Big({\rm C}_\eps^{-1}\cdot{\rm C}_\eps^{\{2\}}\Big)_{11} ,
\end{equation}
where $R_\eps^2$ can have either sign. We then obtain 
\begin{equation} 
\begin{split}
{\cal P}[\eps_g,\eps_{g,\delta}] = \Bigg(\prod_{\vec{x}}\frac{1}{2\pi\sqrt{\det{\rm C}_\eps}}\Bigg) 
\exp\bigg(&{-\frac{1}{2}}\int\dif^3x\,\vec{\eps}(\vec{x})\cdot{\rm C}_\eps^{-1}\cdot\vec{\eps}(\vec{x}) \\
&\! - \frac{1}{2}\frac{R_\eps^2}{P_{\eps_g}^{\{0\}}} \int\dif^3x\, \eps_g(\vec{x}) \nabla^2 \eps_g(\vec{x})\bigg)\,\,.
\end{split} 
\end{equation} 
We can now integrate out $\eps_g$ and $\eps_{g,\delta}$ as before. Rewriting $\eps_g$ 
in terms of the other fields via \eq{pre_main-01} gives, again at leading order in $R_\eps^2$, a contribution of the form 
\begin{equation} 
\begin{split}
\label{eq:pre_main-04-HD}
{\cal P}[\delta_g|\delta] = \Bigg(\!\prod_{\vec{x}}\frac{1}{\sqrt{2\pi P_\eps(\vec{x})}}\!\Bigg)\exp\Bigg[&
	{-\frac{1}{2}}\int\dif^3x\,\frac{\big(\delta_g(\vec{x})-\delta_{g,{\rm det}}(\vec{x})\big)^2}{P_\eps(\vec{x})} \\ 
	&\! -\frac{1}{2}\frac{R_\eps^2}{P_{\eps_g}^{\{0\}}}\!\int\dif^3x\,\big(\delta_g(\vec{x})-\delta_{g,{\rm det}}(\vec{x})\big)
	\nabla^2\big(\delta_g(\vec{x})-\delta_{g,{\rm det}}(\vec{x})\big) 
	\Bigg]\,\,.
\end{split} 
\end{equation} 
The term in the second line can be evaluated straightforwardly 
in Fourier space, and it corresponds to the leading higher-derivative 
stochastic contribution when expanding the result of \cite{Schmidt:2018bkr} 
at first order in $k^2$. There will be other contributions in addition to this one, e.g.~of the form 
(dropping an overall dimensionless coefficient) 
\begin{equation}
\label{eq:additional}
\wp[\delta_g|\delta] 
	\supset\frac{R_\eps^2}{P_{\eps_g}^{\{0\}}}\int\dif^3x\,\delta^n(\vec{x})\, 
	\big(\delta_g(\vec{x})-\delta_{g,{\rm det}}(\vec{x})\big)
	\nabla^2\big(\delta_g(\vec{x})-\delta_{g,{\rm det}}(\vec{x})\big) 
\,\,. 
\end{equation} 
These are less relevant on large scales than the one in \eq{pre_main-04-HD}.

\subsection{About the numerical implementation and the field-dependent covariance}
\label{subsec:numerical_and_renormalization}

\noindent The result of this paper allows for an incorporation of the stochasticities of bias coefficients 
in EFT-based approaches to Bayesian forward modeling \cite{Schmidt:2018bkr,Elsner:2019rql}. 
Without the resummation of these corrections at all orders in the matter field $\delta$ this would not have been possible. 
Indeed, a perturbative calculation would only include them via the Edgeworth-like expansion\footnote{More precisely, 
the loop expansion employed in \cite{Cabass:2019lqx} leads to a functional Taylor series of the logarithm of the likelihood.} 
of \eq{conclusions-02-a} that is not normalizable and hence not amenable to numerical sampling techniques. 
This is in contrast to the likelihood of \eqsII{pre_main-04}{main-13}, 
which is a properly normalized Gaussian with $\delta$-dependent covariance 
(that can be sampled straightforwardly) and reduces to a Dirac delta functional in the limit of vanishing noise amplitudes. 

The Gaussian form of the conditional likelihood, however, is only sensible if we are sure that 
the covariance is positive-definite. Is $P_\eps[\delta](\vec{x})$ a positive number? 
The perturbative analysis of the previous section ensures that the answer is yes if we restrict ourselves to scales 
where the EFT is under control, since the corrections proportional to $\smash{P^{\{0\}}_{\eps_g\eps_{g,{O}}}}$ and 
$\smash{\smash{P^{\{0\}}_{\eps_{g,{O}}\eps_{g,{O}'}}}}$ carry additional powers of $\delta$. 
Notice that the same perturbative arguments apply even in the simple case of only 
Gaussian, scale-dependent noise $\varepsilon_g$ discussed in \cite{Schmidt:2018bkr,Elsner:2019rql,Cabass:2019lqx} 
(see also \cite{2020arXiv200406707S} for a more recent implementation). 
In that case, if we implement the scale dependence of $\smash{P_{\varepsilon_g}(k)}$ via its 
local expansion in powers of $k^2$, we must restrict to scales such that these higher-derivative 
corrections are subleading with respect to the constant part $\smash{P_{\varepsilon_g}^{\{0\}}}$. 

It would nevertheless be nice to show the positivity of the covariance 
non-perturbatively. In order to do this let us consider the manifestly nonnegative combination 
\begin{equation}
\label{eq:conclusions-04}
\bigg(\eps_g(\vec{x})+\sum_{O}\eps_{g,{O}}(\vec{x})\,{O}[\delta](\vec{x})\bigg)^2\,\,. 
\end{equation} 
If we average this combination\footnote{Technically, we multiply it by the joint likelihood 
${\cal P}[\eps_g,\eps_{g,\delta},\dots]$ for the noise fields (given by \eq{pre_main-02} in the 
case of $O=\delta$ only) and functionally integrate over all noise fields.} 
over the noise fields $\eps_g$ and $\eps_{g,O}$, using the fact that they are uncorrelated 
with the matter field $\delta$, we obtain exactly \eq{main-14} times an irrelevant factor proportional 
to a real-space $\smash{\delta^{(3)}_{\rm D}(\vec{0})}$. We then conclude that the 
field-dependent covariance that appears in our likelihood is positive-definite at all orders in perturbations. 

Once we know that the covariance is positive-definite, \eq{pre_main-04} can be straightforwardly built into 
the framework described in \cite{Elsner:2019rql}. The main difference with the implementation presented there is 
that the likelihood is now evaluated by summing over the grid on which $\delta_g$, $\delta_{g,\rm det}[\delta]$ are discretized 
in real space, after applying a sharp-$k$ filter on the scale $\Lambda$ to both fields. 
Work towards implementation of these new terms is currently in progress. We expect that the inclusion of the 
field-dependent covariance will bring perturbative improvements on, e.g., the determination of $\sigma_8$ from 
rest-frame halo catalogs (to cite the observable studied in \cite{Elsner:2019rql} as an example). We expect this 
because in the recent work \cite{2020arXiv200406707S} we have shown this to hold 
for higher-order terms in the deterministic bias expansion, which are more relevant than the ones in the 
field-dependent covariance on large scales. 

Before proceeding, it is worth to emphasize again that, 
from the point of view of the EFTofLSS, the likelihood derived here is 
not complete, in the sense that we have tuned infinitely many counterterms in order to arrive at \eqsII{pre_main-04}{main-13}. 
Had we not done this, we would have obtained many more contributions to the likelihood, among which there 
are those that make it non-Gaussian (see \eq{conclusions-02-b}, for example). 
While these contributions are under perturbative control on large scales, i.e.~when we take the fields appearing 
in the likelihood to contain only long-wavelength modes (as discussed in Sections~\ref{subsec:renormalization} and 
\ref{subsec:perturbative}), they are nevertheless there. It ceases to make sense to neglect them 
but keep the Gaussian likelihood derived here once we go to very high order in the deterministic bias 
expansion, or keep many terms in $P_\eps[\delta]$.

\subsection{Marginalizing over bias parameters}
\label{subsec:marg}

\noindent Ref.~\cite{Elsner:2019rql} showed how the deterministic bias parameters can be 
marginalized over analytically in case of the Fourier-space likelihood with no stochasticities of the $b_O$. 
This becomes very useful once one wants to numerically sample the likelihood via Bayesian methods. 
Here we show the same for the real-space likelihood in \eq{pre_main-04}. 
In the following, we assume that, in preparation, all fields $y \in \{ \d_g, O\}$ are transformed to Fourier space, where we set 
(cf.~the discussion in Section~\ref{subsec:renormalization}) 
\be
y(\vk) = 0\ \forall\ \{ \vk=\vec{0}, |\vk| \geq \Lambda \}\,\,. 
\ee
Then, the fields are transformed back to real space. 

Let us rewrite \eq{main-06} as 
\begin{equation}
\label{eq:mu_definition} 
	\d_{g,{\rm det}}(\vx) = \mu(\vx) +
	\sum_{O\in \O_\text{marg}}
	b_O\, O(\vx) \,\, ,\quad
	\mu(\vx) =
	\sum_{O \in \O_\text{all} \setminus \O_\text{marg}}
	b_O\, O(\vx) \,\, ,
\end{equation}
where $\O_\text{marg}$ denotes the subset of operators whose bias 
parameters we wish to marginalize over (we denote the cardinality of 
this set as $n_\text{marg}$), and we will replace the arguments $[\delta](\vec{x})$ with $\vec{x}$ throughout 
this and the next section for notational clarity. We can then write the likelihood \eq{pre_main-04} as 
(we still keep the continuous integral $\int\dif^3x$; in practical applications this turns into a 
sum over the grid on which $\delta_g$, $\delta_{g,\rm det}$ are discretized in real space) 
\begin{equation}
\label{eq:marginalization_formula}
\begin{split}
{\cal P}\big[\delta_g|\delta,\{b_O\}\big] &= \exp\bigg({-\frac{1}{2}}\int{\rm d}^3x\,\ln 2\pi P_{\eps}(\vec{x})\bigg) \\ 
&\;\;\;\;\times\exp\Bigg[{-\frac{1}{2}}\int\dif^3x\,\frac{\big(\d_g(\vec{x})-\mu(\vec{x})\big)^2}{P_\eps(\vec{x})} \\
&\;\;\;\;\hphantom{\times\exp\Bigg[ } +\sum_{O\in{\cal O}_{\rm marg}} b_O 
\int\dif^3x\,\frac{\big(\d_g(\vec{x})-\mu(\vec{x})\big)O(\vec{x})}{P_\eps(\vec{x})} \\
&\;\;\;\;\hphantom{\times\exp\Bigg[ } - \frac{1}{2}\sum_{O,O'\in{\cal O}_{\rm marg}} b_Ob_{O'} 
\int\dif^3x\,\frac{O(\vec{x})O'(\vec{x})}{P_\eps(\vec{x})}\Bigg] 
\,\,,
\end{split}
\end{equation}
where we have added the argument ``$\{b_O\}$'' to make more clear 
that the likelihood for the data $\delta_g$ is conditioned also on the bias parameters. 
Let us assume that the prior imposed on the set of bias parameters to be marginalized over is Gaussian, so that it can be written as
\be
{\cal P}_\text{prior}(b_O: O \in \O_\text{marg}) = \frac{(2\pi)^{{-\frac{n_{\rm marg}}{2}}}}{\sqrt{\det {\rm C}_\text{prior}}}\, 
\exp\Bigg[{-\frac12} \sum_{O,O'\in \O_\text{marg}} (b_O - \bar b_O) ({\rm C}_\text{prior}^{-1})_{OO'} (b_{O'} - \bar b_{O'})\Bigg]\,\,, 
\ee
where $\bar b_O$ denotes the central value of the prior on the parameter $b_O$ and 
$({\rm C}_\text{prior})_{OO'}$ ($({\rm C}_\text{prior}^{-1})_{OO'}$) denotes the (inverse) covariance. 

Including the prior, \eq{marginalization_formula} can be more compactly written as
\begin{equation}
\label{eq:Ppremarg}
\begin{split}
{\cal P}\big[\d_g|\d, \{b_O\}\big] &= \frac{(2\pi)^{{-\frac{n_{\rm marg}}{2}}}}{\sqrt{\det {\rm C}_\text{prior}}}\, 
\exp\Bigg[{-\frac12} \sum_{O,O'\in \O_\text{marg}} \bar b_O ({\rm C}_\text{prior}^{-1})_{OO'}\bar b_{O'} 
- \frac{1}{2}\int{\rm d}^3x\,\ln 2\pi P_{\eps}(\vec{x})\Bigg] \\
&\;\;\;\;\hphantom{\frac{(2\pi)^{{-\frac{n_{\rm marg}}{2}}}}{\sqrt{\det {\rm C}_\text{prior}}}\, } 
\;\,\times\exp\Bigg[{-\frac{1}{2}} C + \sum_{O\in \O_\text{marg}} b_O B_O 
- \frac{1}{2}\sum_{{O,O'\in \O_\text{marg}}} b_O b_{O'} A_{OO'}\Bigg]\,\,, 
\end{split}
\end{equation}
where
\begin{subequations}
\label{eq:margdefs}
\begin{align}
C &= \int{\rm d}^3x\,\frac{1}{P_\eps(\vx)} \big(\d_g(\vx) - \mu(\vx)\big)^2\,\,, \label{eq:margdefs-1} \\
B_O &= \int{\rm d}^3x\,\frac{\big(\d_g(\vx) -\mu(\vx)\big) O(\vx)}{P_\eps(\vx)} + \sum_{O' \in \O_\text{marg}} 
({\rm C}_\text{prior}^{-1})_{OO'} \bar b_{O'}\,\,, \label{eq:margdefs-2} \\
A_{OO'} &= \int{\rm d}^3x\,\frac{O(\vx) O'(\vx)}{P_\eps(\vx)} 
+ ({\rm C}_\text{prior}^{-1})_{OO'}\,\,. \label{eq:margdefs-3}
\end{align}
\end{subequations}
Note that $A_{OO'}$ is a Hermitian and positive-definite matrix. 
The former is obvious from its definition. The latter follows from the fact that the field-dependent 
covariance is strictly positive, so that the integral $\int{\rm d^3}x\,1/P_\eps(\vec{x})$ defines a scalar product, 
and the fact that the operators $O$ are linearly independent. 
\eq{Ppremarg} then allows us to perform the Gaussian integral over the $b_O$. The result is 
\begin{equation}
\label{eq:Pmarg1}
\begin{split}
{\cal P}\big[\d_g |\d, \{b_O\}_\text{unmarg} \big] &= \Bigg(\prod_{O\in \O_\text{marg}} \int \dif b_O\Bigg)\,
{\cal P} \big[\d_g | \d, \{b_O\} \big] \\
&= \frac{1}{\sqrt{\det {\rm C}_{\rm prior}\det A}}\, 
\exp\bigg[{-\frac{1}{2}}\int{\rm d}^3x\,\ln 2\pi P_{\eps}(\vec{x})\bigg] \\ 
&\;\;\;\;\hphantom{\frac{1}{\sqrt{\det {\rm C}_{\rm prior}\det A}}\, } 
\;\,\times\exp\Bigg[{-\frac{1}{2}} C 
+ \frac{1}{2}\sum_{O,O'\in \O_\text{marg}} B_O(A^{-1})_{O O'}B_{O'}\Bigg]\,\,, 
\end{split}
\end{equation}
where, as it is clear from \eqsIII{mu_definition}{margdefs-1}{margdefs-2}, $C$ and $B_O$ depend only on $\delta$ and 
on the bias parameters that we have not marginalized over. 

We have thus reduced the parameter space from $\{b_O\}$ to 
$\{b_O\}_\text{unmarg}$. This marginalization applies whatever the number $n_{\rm marg}$ 
of bias coefficients to be marginalized over. Notice that $A_{OO'}$ depends on the parameters 
entering the variance $P_\eps(\vx)$, \eq{pre_main-05}, and thus has to be recomputed when those change.

\section*{Acknowledgements}

\noindent We thank Marcel Schmittfull, Ravi Sheth, Marko Simonovi{\'c} 
and Matias Zaldarriaga for useful discussions on related topics.~We 
acknowledge support from the Starting Grant (ERC-2015-STG 678652) 
``GrInflaGal'' from the European Research Council.



\clearpage

\bibliographystyle{utphys}
\bibliography{refs}


\end{document}